\documentclass[twocolumn,superscriptaddress,showpacs,showkeys,preprintnumbers,amsmath,amssymb]{revtex4}
\usepackage{graphicx}%
\usepackage{dcolumn}
\usepackage{amsmath}
\usepackage{latexsym}

\begin{document}
\title{\bf Mechanical Failure of a Small and Confined Solid} 
\author{Debasish Chaudhuri and Surajit Sengupta}
\affiliation{Satyendra Nath Bose National Centre for Basic Sciences, 
Block-JD, Sector-III, Salt Lake,
Calcutta - 700098.}

\begin{abstract}
\vskip0.1cm
\hrule
\vskip0.8cm
\noindent
{\bf Abstract:\,}Starting from  a commensurate triangular thin solid strip, confined within two 
hard structureless walls, a stretch along its length introduces a rectangular 
distortion. Beyond a critical strain the solid fails through nucleation of 
``smectic"-like bands. We show using computer simulations and  
simple density functional based arguments, how a solid-smectic transition 
mediates the failure. 
Further, we show that the critical strain introducing failure is 
{\em inversely} proportional to the channel width i.e. thinner strips are 
stronger ! 
\end{abstract}
\maketitle

\section{Introduction}
\label{intro}
Studies of small systems comprising of a few thousand molecules have 
become increasingly 
important with the advent of nano-technology\cite{nanostuff}. 
In these mesoscopic length scales, a priori, there is no reason that mechanical
behavior will be governed by continuum elasticity theory\cite{micrela}. In 
many situations one needs to study the effect of confinement on the 
structural and other properties of materials. Indeed, hard confinement, in 
one or more directions often induce  new interesting properties; for example, 
the rich phase behaviour of quasi 
two-dimensional colloidal solids\cite{Neser,buckled}
confined between two glass plates showing square, triangular and 
``buckled'' crystalline phases and  re-entrant 
surface melting transition\cite{remelt} of colloidal hard spheres 
quite different from the bulk\cite{alzowe,jaster,srnb} behaviour.
\vskip 0.2cm

In an earlier paper \cite{debc} we have shown that a small confined solid in
quasi one-dimension fails through nucleation of ``smectic" like bands. This is
very unlike a  bulk solid, 
strained beyond it's critical limit, failing through the nucleation and growth 
of cracks\cite{griffith,marder,langer}. The interaction of dislocations 
or zones of plastic deformation\cite{langer,loefsted} with the growing 
crack tip determines the failure mechanism. 
Bulk solid show brittle \cite{SWCNT} or ductile\cite{nano-wire} 
failure depending on these interactions. On the other hand 
the two-dimensional confined
solid shows ductile failure along with {\em reversible plasticity} --- the 
plastic stress vanishes when the strain is removed and the fractured
parts join up without any discontinuity. This is because of the high 
amount of orientational order imposed by the confining walls which 
ensure that each smectic band is confined within a dislocation - 
antidislocation pair. On the removal of the strain these defects 
annihilate in pairs leaving a perfect triangular lattice identical to 
the initial solid.  

In section \ref{system} we describe the phase structure of the two dimensional 
confined strip and possible phase transitions. We illustrate our discussion 
with previously published\cite{debc} computer simulation data. 
In section \ref{theory} we present simple density functional\cite{rama,Chaikin}
arguments showing the difference between a bulk solid, confined-solid and a 
smectic phase as well as a  Lindeman-like  criterion\cite{Zahn} for 
nucleation of the smectic. In section \ref{results} we give the results and 
discuss their implications. We end by offering some concluding remarks and 
discussions on future directions of research in section \ref{conclusion}.
\vskip 0.2cm

\section{The system}
\label{system}
\vskip .2 cm
\begin{figure}[t]
\begin{center}
\includegraphics[width=8.0cm]{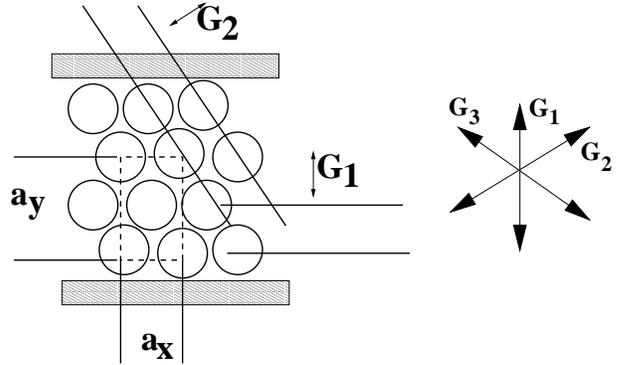}
\end{center}
\caption{The confined solid is shown along with the centered rectangular (CR) unit cell. For 
an unstrained triangular lattice $a_x = a_0$ and $a_y = \sqrt{3} a_0$. ${\bf G_1}$,  ${\bf G_2}$ and ${\bf G_3}$ denote the directions of the three reciprocal lattice vectors (RLV). The third reciprocal lattice direction ${\bf G_3}$ is equivalent to the direction ${\bf G_2}$, even in presence of the walls.
}
\label{wallpic}
\end{figure}

A two dimensional bulk solid made of molecules interacting via spherically 
symmetric potentials always stabilizes\cite{Chaikin} in the triangular 
lattice configuration. 
For specificity and simplicity we consider hard~-disk molecules, which are 
restricted from overlapping with each other due to an infinitely large energy 
cost and remains non-interacting when they do not overlap.
Apart from being easily  accessible to theoretical 
treatment\cite{hamac}, experiments with nearly 
``hard'' (macro-)molecules  viz. sterically stabilized 
colloids\cite{colbook} are possible. The hard~-disk free energy is entirely 
entropic in 
origin and the only thermodynamically relevant variable is the number density   
$\rho = N/V$ or the packing fraction $\eta = (\pi/4) \rho {\rm d}^2$.
The energy scale of the system is entirely set by $k_B T$ where $k_B$ is
the Boltzmann constant and $T$ the (kinetic) temperature. 
Accurate computer simulation\cite{jaster,srnb} vindicated the 
defect unbinding theory\cite{kt} of melting of the 
hard~-disk solid below $\eta_m = .706$.

Imagine a narrow channel in two dimensions  of width $L_y$ defined by 
hard walls at $y = 0$ and $L_y$ ($V_{\rm wall}(y) = 0$ for 
$ {\rm d/2} < y < L_y - {\rm d/2}$ and $ = \infty$ otherwise) and 
length $L_x$ with $L_x \gg L_y$.  
Therefore, $n_l$ layers of an undistorted triangular solid of lattice 
parameter $a_0$ and diameter d may be accommodated 
(Fig.\ref{wallpic}) if $L_y$ is commensurate\cite{debc} 
with the~ solid packing fraction, {\em i.e.}
\begin{equation}
L_y = \frac{\sqrt{3}}{2}(n_{l} - 1) a_0 + {\rm d}.
\label{perfect}
\end{equation}

For a system with a constant number of particles $N$ and $L_y$, $a_0$ is decided
by the packing fraction $\eta$ alone. Note that $L_x=n_x a_0 = N a_0/n_l$, 
and $a_0$ is given by $\rho = N/L_x L_y$. Defining $\chi(\eta, L_y) = 1 + 2(L_y - {\rm d})/\sqrt{3} a_0$,
the above condition reads $\chi = {\rm integer} = n_{l}$. Violation
of Eqn.(\ref{perfect}) induces an {\em internal} rectangular strain measured 
from a reference triangular lattice of $n_l$ layers;
\begin{eqnarray}
\varepsilon_d & = & \varepsilon_{xx} - \varepsilon_{yy} \\ \nonumber
              & = & \frac{a_x-a_0}{a_0} - \frac{a_y - a_0\,\sqrt{3}/2}{a_0\,\sqrt{3}/2}.
\label{strain1}
\end{eqnarray}
The lattice parameters of the strained, centered~-rectangular (CR) unit 
cell $a_x$ and $a_y$ are shown in Fig.\ref{wallpic}. 
\vskip 0.2cm

\vskip .2 cm
\begin{figure}[h]
\begin{center}
\includegraphics[width=7.0cm]{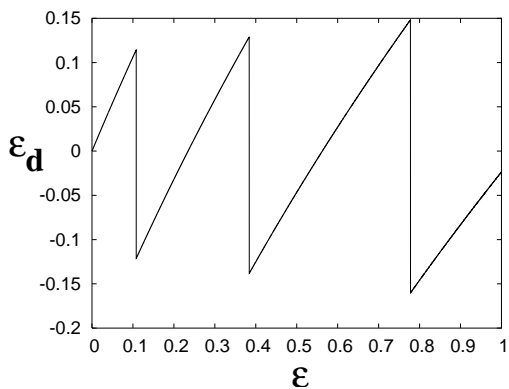}
\end{center}
\caption{ A plot of internal strain $\varepsilon_d$ as a function of external 
strain $\epsilon$. The jumps in $\varepsilon_d$ corresponds to half-integral 
values of $\chi$.
}
\label{epsd}
\end{figure}

The {\em external} strain, on the other hand, imposed by rescaling 
$L_x$ keeping $L_y$ fixed, $\epsilon = (\eta_0-\eta)/\eta$ is measured 
using the initial triangular lattice as the reference state. As the solid is 
free to move locally, it may readjust itself to increase (or decrease) the 
number of layers $n_l$ in response to the external strain ensuring that the 
internal deviatoric strain remains small (minimization of free energy). 
It can be shown that\cite{debc},
\begin{equation}
\varepsilon_d =  \frac{n_l - 1}{\chi - 1} - \frac{\chi - 1}{n_l - 1},
\label{strain}
\end{equation}
where the number of layers $n_l$ is the nearest integer to $\chi$ so that 
$\varepsilon_d$ has a discontinuity at half-integral values of $\chi$. 
 This internal strain $\varepsilon_d$ is related 
non-linearly to $\epsilon$ and may remain small even if $\epsilon$ is large 
(Fig.\ref{epsd}).
\vskip .2 cm
\begin{figure}[h]
\begin{center}
\includegraphics[width=9.0cm]{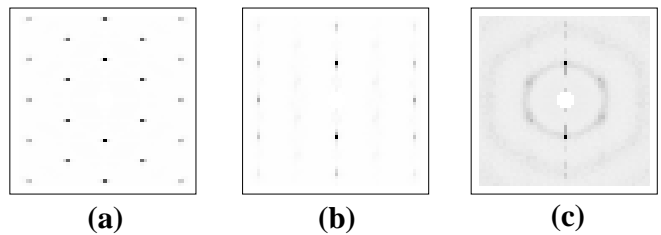}
\end{center}
\caption{Structure factors of different phases. (a), (b) and (c) are the 
structure factors corresponding to solid, smectic and modulated liquid phases 
respectively.
}
\label{sk}
\end{figure}

\vskip .2cm
Walls induce density modulations, with peaks and troughs running parallel 
to the walls. The ensemble average of the local density
$\rho({\bf r}) = <\sum_{i=1,N}\delta({\bf r - r_i})> $  
shows asymmetric (ellipsoidal) density profiles at the lattice points, 
with the semi-major axes lying along the walls. 
The structure factor is defined as
$\rho_{\bf G} =\left| \left< \frac{1}{N^2} \sum_{j,k = 1}^N
\exp(-i {\bf G}.{\bf r}_{jk})\right> \right|.$
For ${\bf G} = \pm {\bf G_1}(\eta)$, the
reciprocal lattice vector (RLV) correspond to the set of close-packed
lattice planes of the CR lattice perpendicular to the
wall, and for ${\bf \pm G_2}(\eta)$ and ${\bf \pm G_3}(\eta)$ the
four equivalent RLVs for close-packed planes at an angle
( $ = \,\pi/3$ and $2\pi/3$ in the triangular lattice) to the wall
(see Fig. \ref{wallpic}). It is useful to define the 
Lindemann parameter
$l = < ({u^x}_i - {u^x}_j)^2>/a_x^2 + < ({u^y}_i - {u^y}_j)^2>/a_y^2 $
where the angular brackets denote averages over configurations,
$i$ and $j$ are nearest neighbors and ${u^{\alpha}}_i$ is the $\alpha$-th
component of the displacement of particle $i$ from it's mean position.
The parameter $l$ diverges at the melting transition \cite{Zahn}.

\begin{figure}[h]
\begin{center}
\includegraphics[width=8.0cm]{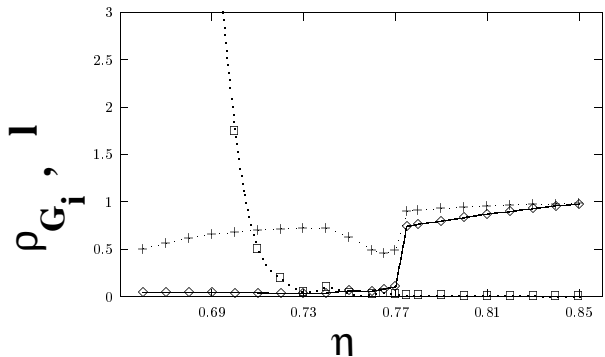}
\end{center}
\caption{Results of Monte Carlo (MC) simulations of
$N = 650$ hard~-disks confined between two
hard walls separated by a distance
$L_y = 9\,{\rm d}$. For each $\eta$, the
system was equilibrated over $10^6$ MC steps (MCS) and data
averaged over a further $10^6$ MCS.
At $\eta = 0.85$ we have a strain free triangular lattice.
Plots show the structure factors $\rho_{\bf G_i}, i = 1 (+),2(\diamond)$
for RLVs ${\bf G_i}(\eta)$, averaged over
symmetry related directions, as a function of $\eta$ and
the Lindemann parameter $l(\Box)$.
The lines in the figure are a guide to the eye. At $\eta \approx .78$
$\rho_{\bf G_2}$ jumps to zero while $\rho_{\bf G_1}$ shows a small 
dip. The region with $.7< \eta < .78$, therefore, contains a smectic
phase which melts for $\eta < .7$, at which point the Lindemann 
parameter $l$ diverges. 
}
\label{order}
\end{figure}

\vskip .2cm
Apart from the solid phase (Fig.\ref{wallpic} and Fig.\ref{sk}(a)), the
externally imposed density modulation may give rise to a modulated liquid or a 
smectic phase. Structurally a smectic is made up of overlapping asymmetric 
local density profiles along the walls generating continuous strips of density 
maxima running parallel to the walls. Peaks in the structure factor 
corresponding to ${\bf \pm G_2}(\eta)$ vanish, although for 
${\bf G} = \pm {\bf G_1}(\eta)$, one continues to obtain strong peaks
Fig.\ref{sk}(b). In this phase, however, inter-layer 
particle exchanges are suppressed causing the Lindemann parameter to remain 
small. In  a modulated liquid, on the other hand, this interlayer 
exchange is large. The structure factor of a modulated liquid 
(Fig.\ref{sk}(c)) features a ring-like maximum characteristic of liquid, in 
addition to somewhat strong peaks corresponding to ${\bf G_1}$. Transitions 
among these phases are observed as the external strain is imposed strating from
a perfect triangular solid. The sequence of phase changes is shown in
Fig.\ref{order}, the result of extensive Monte Carlo simulations of this 
system\cite{debc}.

\vskip .5 cm
 
\section{Theory}
\label{theory}
As shown earlier\cite{debc}, the failure of commensurate solid under 
tensile strain imposed in the manner discussed in the previous section, 
comes about through the nucleation of smectic bands within the solid. 
Monte-Carlo simulations show, at half-integral $\chi$ where the local 
internal strain $\varepsilon_d$ becomes 
discontinuous, $\rho({\bf r})$ at nearest neighbour sites 
overlap along the $x$-direction, parallel to the walls, generating 
smectic bands. The stress associated with $\varepsilon_d$ vanishes at 
these points and the solid fails under tension. In this section we shall 
show, using simple density functional\cite{rama,Chaikin} arguments, that 
the phase transition and the consequent tensile failure (a smectic 
cannot support stress parallel
to the smectic layers) is brought about by this overlap in the local density. 
Since mechanical failure in our system is a consequence of a phase transition, 
it is reversible --- as the strain is reduced back to zero, the stress also 
vanishes and the perfect triangular lattice is recovered\cite{debc}. 

Within density functional theory\cite{rama}, the excess grand potential of a non-uniform 
liquid containing a density modulation $\rho({\bf r})$ over the uniform
liquid of density $\rho_l$ is given by,

\begin{eqnarray}
\frac{\Delta \Omega}{k_B T} & = &\int d{\bf r}[\rho({\bf r})\log(\rho({\bf r})/\rho_l)-\delta\rho({\bf r})] \nonumber \\
                           &   & -\frac{1}{2}\int d{\bf r'}C(|{\bf r - r'}|)\delta\rho({\bf r})\delta\rho({\bf r'}). 
\label{rky}
\end{eqnarray}

Here $\delta\rho({\bf r}) =\rho({\bf r})-\rho_l$ and $C(r)$ is the direct 
correlation function of the uniform liquid\cite{hamac}. A functional 
minimization of the free energy yeilds the following self-consistency 
equation for the density:

\begin{equation}
\frac{\rho({\bf r})}{\rho_l} = \exp[\int d{\bf r'}C(|{\bf r - r'}|)
\delta\rho({\bf r'})]
\end{equation}

In principle one should solve the above equation within the constraints
imposed by the walls and obtain the equilibrium $\rho({\bf r})$. Substitution
of this $\rho({\bf r})$ into Eqn.\ref{rky} gives the equilibrium free energy
and phase transitions. While we intend to carry out this procedure in the 
future, we must point out that for the present problem, this is complicated
by surface terms and anisotropic, external, fields which are difficult to 
incorporate. In this paper we shall take a much simpler route in exploring
the various conditions for the solid~-smectic transition given the nature 
of the $\rho_{\bf G_i}$ (the order parameters) obtained from our simulations.  

One may expand, therefore, the logarithm of the local density 
profile $\log\rho({\bf r})$ 
in a Fourier series\cite{Chaikin} around a lattice point at the origin, 
to get, 
\begin{equation}
\rho({\bf r}) = {\cal N} \exp\left(2 C_0\sum_{i=1}^3 \rho_{\bf G_i}
\cos ({\bf G_i.r} ) \right)
\label{2nd}
\end{equation}
where $C_0$ is a constant, of order unity, denoting the Fourier transform of 
the direct correlation function calculated at a q-vector corresponding to 
the smallest RLV set of the solid. We have kept contributions only from 
this set. 

For a perfect triangular lattice, the RLV's are 
${\bf G_1} = \hat{y} \frac{2\pi}{d_y}$, ${\bf G_2} = \hat{x} \frac{2\pi}{d_y}\cos(\frac{\pi}{6}) + \hat{y} \frac{2\pi}{d_y}\sin(\frac{\pi}{6})$ and ${\bf G_3} = \hat{x} \frac{2\pi}{d_y}\cos(\frac{\pi}{6}) - \hat{y} \frac{2\pi}{d_y}\sin(\frac{\pi}{6})$, 
where $d_y = \frac{\sqrt{3}}{2}a_0$. Using these relations and the fact that 
in the presence of confining walls the Fourier amplitudes denoting solid 
order are virtually constant upto the transition and 
$\rho_{\bf G_2} = \rho_{\bf G_3} \ne \rho_{\bf G_1} $, 
Eqn.\ref{2nd} gives,
\begin{widetext}
\begin{equation}
\rho({\bf r}) = {\cal N} \exp\{C_0 (2 \rho_{\bf G_1} + 4 \rho_{\bf G_2})\}
\exp\left(-\frac{1}{2} C_0 \left(\frac{2\pi}{d_y}\right)^2\left\{ (2 \rho_{\bf G_1} + \rho_{\bf G_2})y^2 + 3 \rho_{\bf G_2} x^2 \right\}\right)
\label{3rd}
\end{equation}
\end{widetext}
Clearly the density profile is Gaussian, of the form, 
$\rho({\bf r}) \sim exp(-y^2/2\sigma_y^2 - x^2/2\sigma_x^2)$. 
Therefore, the spreads of density profile in $x$ and $y$-directions are 
given by $\sigma_x$ and $\sigma_y$ respectively, with
\begin{eqnarray}
\sigma_x^2 &=& \frac{1}{C_0} \left(\frac{d_y}{2\pi}\right)^2 
\frac{1}{3 \rho_{\bf G_2} }\\
\sigma_y^2 &=& \frac{1}{C_0} \left(\frac{d_y}{2\pi}\right)^2 
\frac{1}{2 \rho_{\bf G_1} + \rho_{\bf G_2}}.
\label{4th}
\end{eqnarray}

In the absence of walls, $\rho_{\bf G_1} = \rho_{\bf G_2}$ making 
$\sigma_x = \sigma_y$, {\em i.e.} the density profile comes out to be 
symmetric in both directions, as expected for the bulk triangular solid. 
The presence of 
walls make $\sigma_y < \sigma_x$ making the density profile elliptical with 
larger spread in $x$-direction, the direction parallel to the walls. Two 
neighbouring density profiles will overlap to form a smectic if 
$\sigma_x > a_x$. This leads us to the definition of a measure of overlap 
${\cal O}_l = \left(\sigma_x/a_x \right)$.  The condition ${\cal O}_l > 1$ 
is then the Lindemann criterion for nucleation of the smectic phase. 
Remembering $a_x = a_0 (n_l - 1)/(\chi -1)$, we get,
\begin{equation}
{\cal O}_l = \frac{1}{4\pi} \frac{1}{\sqrt{C_0\rho_{\bf G_2}}} \frac{\chi -1}{n_l -1}. 
\end{equation}

Whenever, $\rho_{\bf G_2}\to 0$ {\em i.e.} with the loss of solid order 
${\cal O}_l$ diverges although $\sigma_y$ remains finite, since 
$\rho_{\bf G_1} \neq 0$ in presence of the walls. This indicates a 
solid-smectic transition. However, even before $\rho_{\bf G_2}\to 0$ the 
quantity 
 $\Delta = \frac{\chi -1}{n_l -1}$ and therefore ${\cal O}_l$ shows large 
jumps at those internal strain ($\varepsilon_d$) values where $\chi$ becomes
 half-integer. It is interesting to note that, at these points 
$\varepsilon_d$ has discontinuities and the system fails\cite{debc}. This 
shows that the mode of failure predicted by our theory is 
through a solid-smectic transition. The fact that $\rho_{\bf G_1} $ remains 
non-zero even at very small densities, due to the confinement from the walls, 
gives rise to the density modulations in the confined liquid.

We have shown therefore that the overlap in the density profiles may be used 
as an ``order parameter'' for the solid to smectic transition. We show below
that jumps in this order parameter tantamounts to mechanical failure of the 
solid.

\section{Results and Discussion}

\label{results}
\begin{figure}[t]
\begin{center}
\includegraphics[width=8.0cm]{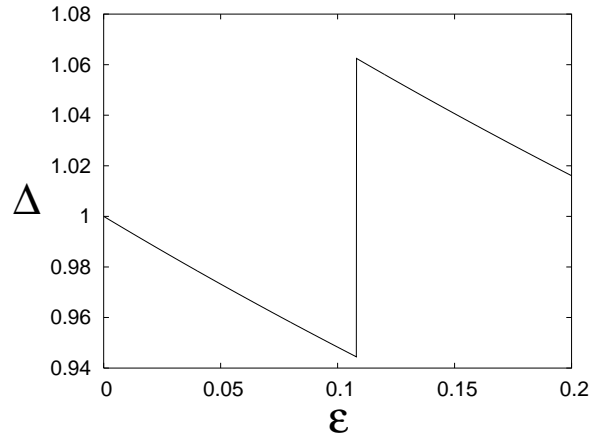}
\end{center}
\caption{For a 10-layered solid with $L_y$ commensurate with the initial strainless triangular structure at $\eta=.85$ overlap term $\Delta$ is plotted as a function of external strain $\epsilon$. Density profile overlap shows a jump increase at strains $\epsilon > .1$, the failure strain value\cite{debc}. 
}
\label{epsD}
\end{figure}

We begin with studying the overlap $\Delta$ as a function of external 
strain $\epsilon$. For specificity, we start from a triangular solid of 
packing fraction $\eta = .85$ with $L_y$ commensurate with a $n_l = 10$ 
layered solid. With increasing strain initially the overlap $\Delta$ reduces 
due to increased separation ($a_x$) between neighbouring lattice points. 
But above a strain ($\epsilon$) of about $10\%$, $\chi$ reaches the 
half-integral mark and $\Delta$ shows a discontinuous increase, indicating 
large 
overlap between neighbouring density profiles along the wall ; indicating a 
solid to smectic transition (Fig.\ref{epsD}) at
the  failure strain $\epsilon^\ast$. With further 
increase in strain the overlap reduces, again due to increased separation 
between neighbouring lattice points. At higher strains the smectic melts 
into a modulated liquid due to increased fluctuations connected with the 
reduced density\cite{debc}.

\begin{figure}[t]
\begin{center}
\includegraphics[width=8.6cm]{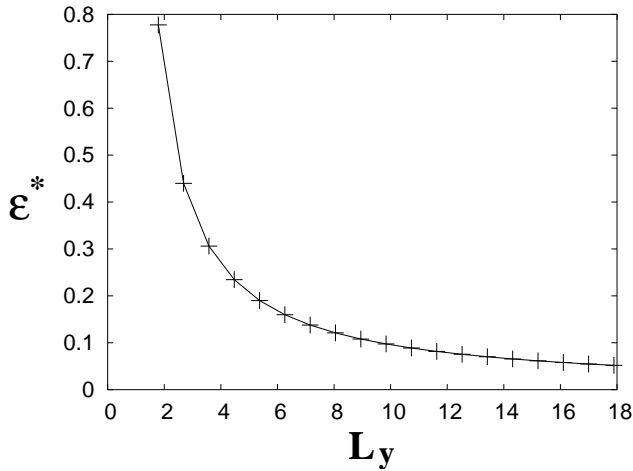}
\end{center}
\caption{Failure strains $\epsilon^\ast$ for various interwall 
separations $L_y$ confining $n_l=2 \to 20$ layered triangular strips
 at $\eta = .85$ is plotted. Failure strain decreases with increase 
in $L_y$.
}
\label{fails}
\end{figure}

We have performed this calculation for various $L_y$ values commensurate 
with starting
triangular solids of $n_l = 2, 3 \dots 20$ layers at packing fraction 
$\eta=.85$. We found out the failure strains $\epsilon^\ast$ at each $L_y$ 
and plotted them in Fig.\ref{fails} as a function of $L_y$. This clearly 
shows that the failure strain reduces with increase in $L_y$. This 
demonstrates the fact, derived earlier from Monte-Carlo 
simulations\cite{debc}, that thinner (smaller $L_y$) strips are 
stronger!
 
\begin{figure}[h]
\begin{center}
\includegraphics[width=8.6cm]{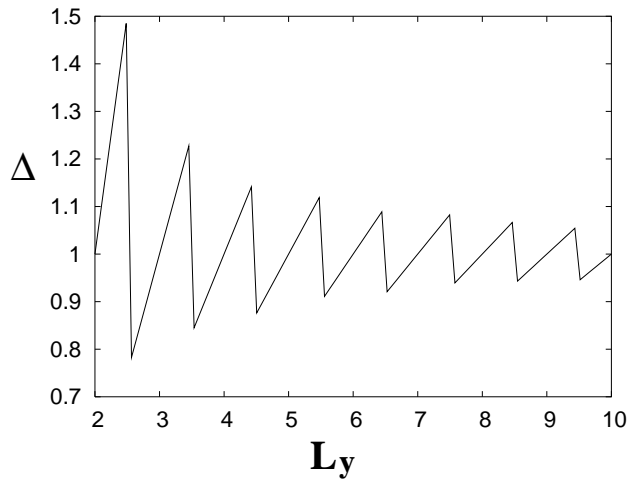}
\end{center}
\caption{Overlap $\Delta$ is plotted for a system at $\eta = .85$ with changing interwall separation $L_y$. Amount of smectic overlaps $\Delta$ at failures reduces with increasing $L_y$.
}
\label{ly}
\end{figure}

In Fig.\ref{ly} we have plotted the overlap term $\Delta$ with increasing 
interwall separation $L_y$ at $\eta = .85$. The jumps, as usual, indicate 
failure strains corresponding to discontinuities in the internal 
strain $\varepsilon_d$
at half-integral values of $\chi$. The plot shows that the amount of 
overlap at the failure strains $\epsilon^\ast$ reduces with increasing  
$L_y$ indicating that at large interwall separations the system starts 
to behave as a bulk solid and more conventional modes of failure {\em viz.} 
through formation and interaction of cracks and twin boundaries starts 
becoming active. 

\section{Conclusion}
\label{conclusion}

In this paper we have described how mechanical and phase behaviour 
are intimately related to each other for a narrow solid strip confined 
between hard walls. Both may therefore be described using 
simple arguments based on density functional theory. The particles of the 
solid are assumed to interact via hard~-disk interactions 
Apart from constrained hard~-sphere colloids\cite{colbook} where our results 
are directly testable, a similar fracture mechanism may be 
observable in experiments on the deformation of mono-layer nano beams 
or strips of real materials provided the confining channel is made of a 
material which is harder and has a much smaller atomic size than that of
the strip\cite{nanostuff}. 

We believe that phase transition in confined systems, as described in this 
paper, may have future applications in nano-technology. 
The feature of reversible plasticity\cite{debc} 
may be used to produce nano scale mechanical switches. 
For example, thermal conductivity and electrical conductivity  of such 
narrow strips after nucleation of smectic bands are expected to reduce 
drastically. These properties need to be investigated in detail. Due to 
reversible plasticity a small change in strain in opposite directions can 
vary these properties of confined nano-strips giving rise to a possible use
 of it as mechanical switch in nano-devices.
\vskip .2 cm

\section{Acknowledgement}
\label{ackno}
The authors thank M. Rao and V. B. Shenoy  
for useful discussions; D. C. thanks C.S.I.R., India, for a fellowship. 
Financial support from DST grant SP/S2/M-20/2001
is gratefully acknowledged.

 


\begin{thebibliography}{99}
\bibitem{nanostuff}
R. Giao {\em et. al.}, \prl {\bf 85}, 622 (2000);
A. N. Cleland and M. L. Roukes, Appl. Phys. Lett. {\bf 69}, 2653 (1996).
\bibitem{micrela}
I Goldhirsh and C. Goldenberg, European Physical Journal {\bf E 9}, 245 (2002).
\bibitem{Neser}
S. Neser et al. Phys. Rev. Lett. 79, 2348, (1997).
\bibitem{buckled}
M. Schmidt and H. Loewen, Phys. Rev. Lett. {\bf 76}, 4552, (1996);
T. CHou and D. R. Nelson, Phys. Rev. E, {\bf 48}, 4611 (1993).
\bibitem{remelt}
R. P. A. Dullens and W. K. Kegel, Phys. Rev. Lett. {\bf 92}, 195702 (2004).
\bibitem{alzowe}
B.J. Alder, T.E. Wainwright, \prb {\bf 127}, 359 (1962);
J.A. Zollweg, G.V. Chester, P.W. Leung, \prb {\bf 39} 9518 (1989);
H. Weber, D. Marx, Europhys. Lett. {\bf 27} 593 (1994).
\bibitem{jaster}
A. Jaster, Physica A.{\bf 277}, 106 (2000). 
\bibitem{srnb}
S. Sengupta, P. Nielaba, K. Binder, Phys. Rev. {\bf E 61}, 6294 (2000).
\bibitem{debc}D. Chaudhuri, S. Sengupta, \prl {\bf 93}, 115702 (2004).
\bibitem{kt}
Kosterlitz J M and Thouless D J 1973 J. Phys. C: Solid State Phys. {\bf 6} 1181, Halperin B I and Nelson D R 1978 \prl {\bf 41} 121, Nelson D R and Halperin B I 1979 \prb {\bf 19} 2457, Young A P 1979 \prb {\bf 19} 1855.
\bibitem{marder}
J. A. Hauch, D. Holland, M. P. Marder, and H. L. Swinney,
 Phys. Rev. Lett. {\bf 82}, 3823 (1999). D. Holland and M. Marder, 
Phys. Rev. Lett. {\bf 80}, 746 (1998)
\bibitem{langer}
J. S. Langer, Phys. Rev. E {\bf 62}, 1351 (2000).
\bibitem{griffith}
A. A. Griffith, Philos. Trans. Roy. Soc. {\bf A 221}, 163 (1920).
\bibitem{loefsted}
R. L\"ofstedt, Phys. Rev. E {\bf 55}, 6726 (1997).
\bibitem{SWCNT}
T. Belytschko {\em et. al.}, Phys. Rev. B {\bf 65}, 235430 (2002);
M. F. Yu {\em et. al.}, Science {\bf 287}, 637 (2000).
\bibitem{nano-wire}
H. Ikeda {\em et. al.}, Phys. Rev. Lett. {\bf 82}, 2900 (1999);
P. S. Branicio and J.-P. Rino, Phys. Rev. B {\bf 62}, 16950 (2000).
\bibitem{rama}
T. V. Ramakrishnan and M. Yussouff, Phys. Rev. B {\bf 19}, 2775 (1979).
\bibitem{Chaikin} P. M. Chaikin and T. C. Lubensky {\em Principles of 
condensed matter physics}, (Cambridge University Press, 1995).
\bibitem{Zahn}
K. Zahn, R. Lenke and G. Maret, Phys. Rev. Lett. {\bf 82}, 2721 (1999)
\bibitem{hamac}
J. P. Hansen and I. R. MacDonald {\em Theory of simple liquids} 
(Wiley, Cluchester, 1989).
\bibitem{colbook}
I. W. Hamley {\em Introduction to Soft Matter: polymer, colloids, amphiphiles
and liquid crystals} (Wiley, Cluchester, 2000).

\end{thebibliography}
\end{document}